\documentclass[aps,pre,twocolumn]{revtex4}
\usepackage{amsfonts,amssymb,amsmath,bm}
\usepackage{graphicx,epsfig,color}
\usepackage{citesort}

\begin{document}

\title{Coherent structures and extreme events in rotating multiphase turbulent flows}

\author{L. Biferale}
\affiliation{Department of Physics and INFN, University of Rome Tor Vergata, Via della Ricerca Scientifica 1, 00133 Rome Italy}
\author{F. Bonaccorso}
\affiliation{Department of Physics and INFN, University of Rome Tor Vergata, Via della Ricerca Scientifica 1, 00133 Rome Italy}
\author{I.M.  Mazzitelli} 
\affiliation{Department of Physics and INFN, University of Rome Tor Vergata, Via della Ricerca Scientifica 1, 00133 Rome Italy}
\author{M.A.T. van Hinsberg} \affiliation{Department of Physics,
  Eindhoven University of Technology, 5600 MB Eindhoven, 
  Netherlands }
\author{A.S. Lanotte} \affiliation{CNR-ISAC and INFN,
  Strada Provinciale Lecce-Monteroni, 73100 Lecce, Italy}
\author{S. Musacchio} \affiliation{Universit\'e de Nice Sophia
  Antipolis, CNRS, Laboratoire J. A. Dieudonn\'e, UMR 7351, 06100
  Nice, France}
\author{P. Perlekar}
\affiliation{TIFR Centre for Interdisciplinary Sciences, 21 Brundavan Colony, Narsingi, Hyderabad 500075, India}
\author{F. Toschi} \affiliation{Department of Physics, Eindhoven
  University of Technology, 5600 MB Eindhoven,  Netherlands and IAC CNR, Via dei Taurini 19, 00185 Roma, Italy}

\begin{abstract}
{\bf Post print version of the paper published on Phys. Rev. X {\bf 6} 041036 (2016) DOI: 10.1103/PhysRevX.6.041036}\\
\noindent By using direct numerical simulations (DNS) at unprecedented resolution we
study turbulence under rotation in the presence of simultaneous direct and
inverse cascades. The accumulation of energy at large scale leads to
the formation of vertical coherent regions with high vorticity
oriented along the rotation axis.  By seeding the flow with millions
of inertial particles, we quantify -for the first time- the effects of
those coherent vertical structures on the preferential concentration
of light and heavy particles.  Furthermore, we quantitatively show
that extreme fluctuations, leading to deviations from a
normal-distributed statistics, result from the entangled interaction
of the vertical structures with the turbulent background.  Finally, we
present the first-ever measurement of the relative importance
between Stokes drag, Coriolis force and centripetal forces along the
trajectories of inertial particles.  We discover that vortical
coherent structures lead to unexpected diffusion properties for heavy
and light particles in the directions parallel and perpendicular to
the rotation axis.
\end{abstract}

\maketitle 
\section{INTRODUCTION}
The dynamics of fluids under strong rotation is a challenging problem
in the field of hydrodynamics and magnetohydrodynamics
\cite{G68,Da13}, with key applications to geophysical and
astrophysical problems (oceans, Earth's atmosphere and inner mantle,
gaseous planets, planetesimal formations) and engineering
(turbomachinery, chemical mixers) \cite{Ba01,Ch08,Du04,L83,HB98,O97}.
A considerable number of experiments
\cite{HBG82,SDD08,MMRS11,D11,YVS13,GCCM14,YZ98,SW99,TM09,YMK11,TM11,CCEH05,SMRP12,Al15}
have been devoted to investigating how turbulence is affected by rotation
(for a recent review of experimental and numerical results see Ref.
\cite{GM15}).
\begin{figure}[h!]
\centering
\includegraphics[angle=270,width=1.0\columnwidth]{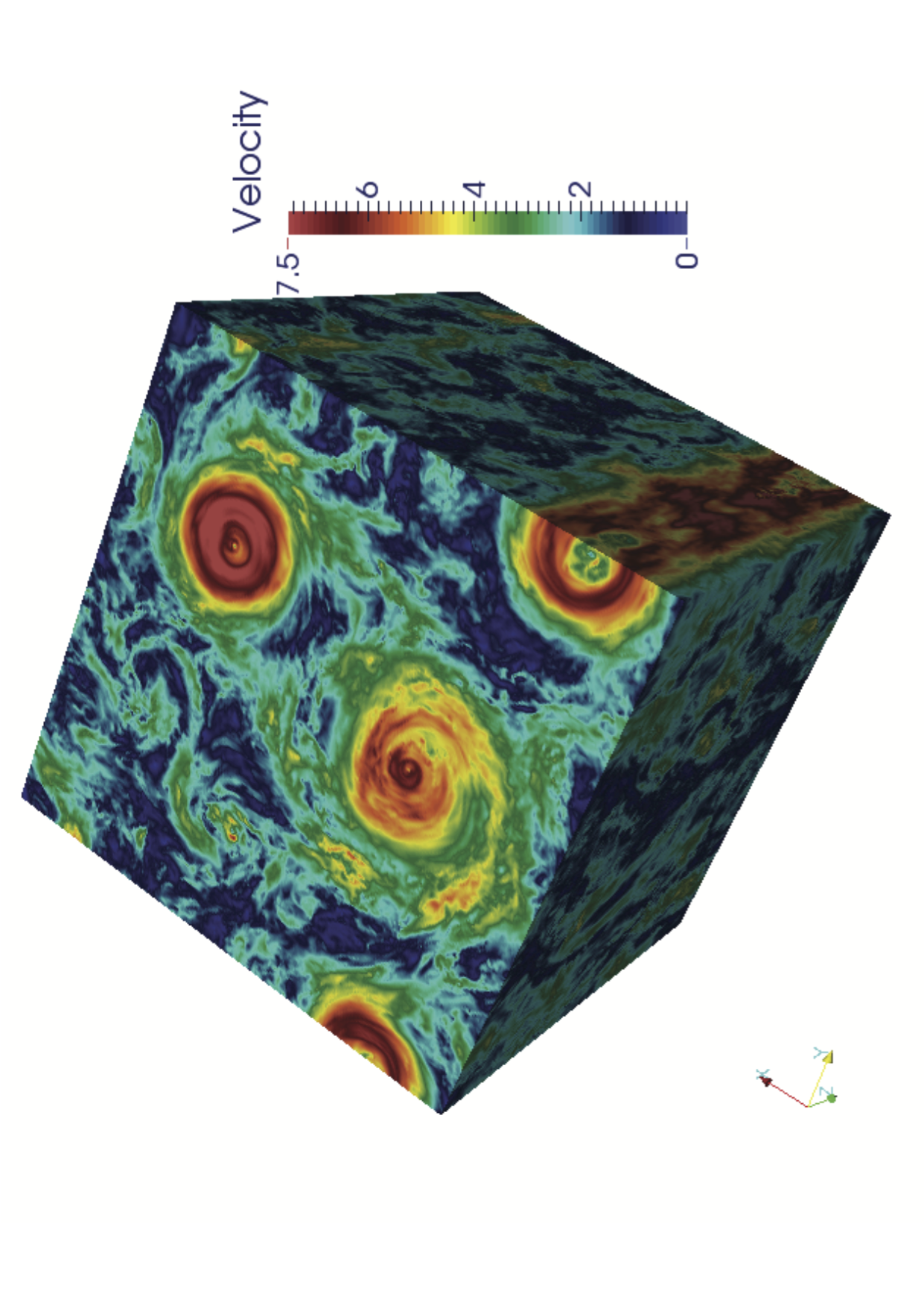}
\includegraphics[angle=270,width=1.0\columnwidth]{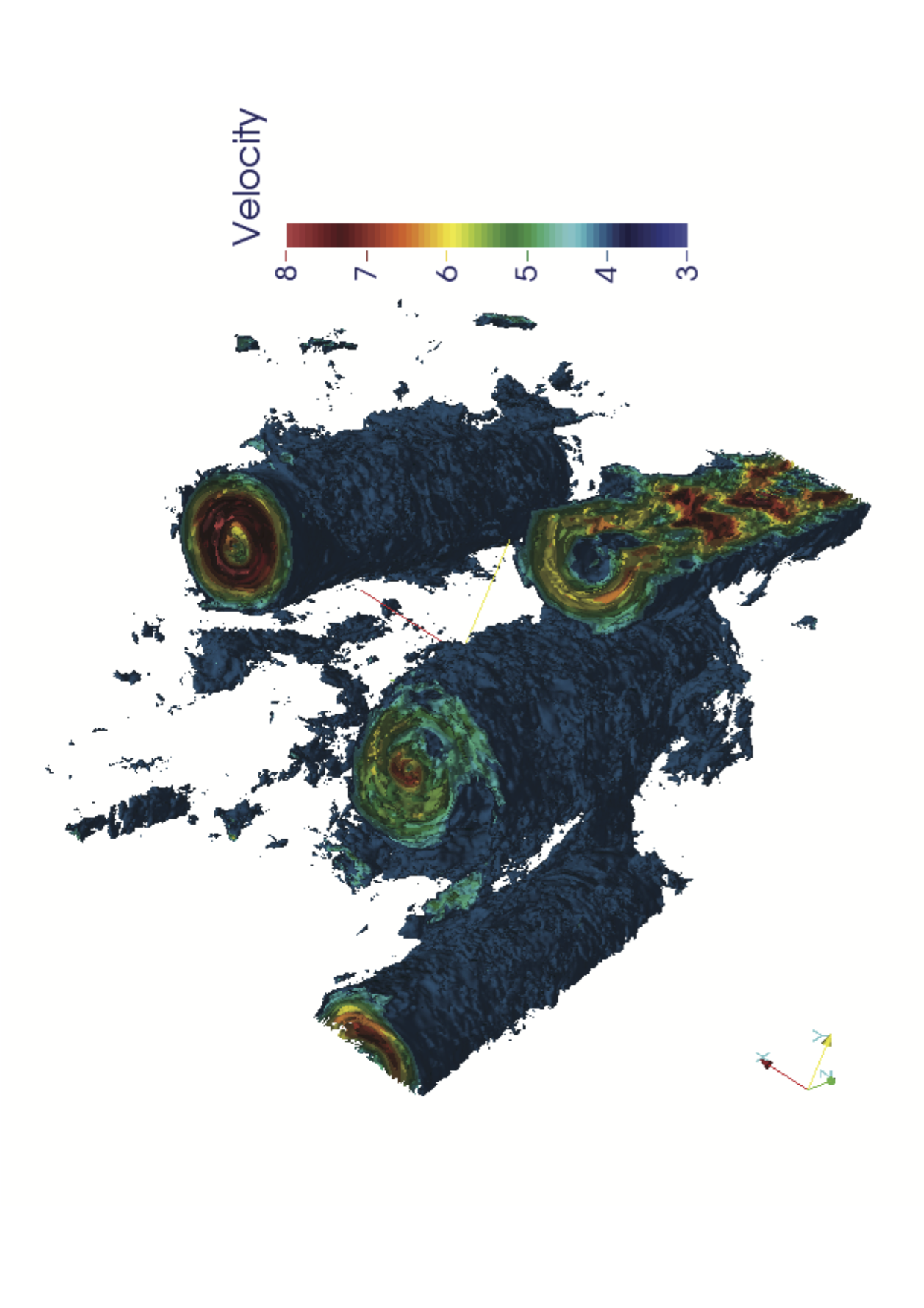}
\caption{Top: a 3D rendering of a turbulent flow at Rossby number $Ro
  =0.25$ and rotation rate $\Omega=10$. An inverse energy cascade is
  present in the turbulent dynamics. The stationary
  behavior is characterized by the formation of three cyclonic
  coherent columnar vortices emerging from the background of 3D
  turbulent fluctuations. Bottom: vortical
  structures parallel to the rotation axis. Note the turbulent
  fluctuations exist also inside the core of each vortex. Color
   scale is based on  the velocity amplitude.}
\label{fig:1}
\end{figure}
The strength of rotation is measured by the Rossby number
$Ro=(\epsilon_f k_f^2)^{1/3}/\Omega$, defined as the ratio of the
rotation time, $\tau_\Omega = 1/\Omega$, and the flow time-scale,
$\epsilon_f k_f^2$. Here $\epsilon_f$ and $k_f$ are the input of
energy and the wavenumber where the external forcing is applied (see
table 1). The most striking phenomenon originated by the Coriolis
force is the formation of intense and coherent columnar vortical
structures (see Fig. \ref{fig:1}), which has been observed in
numerical simulations \cite{YZ98, SW99, TM09, YMK11, TM11} and in
experiments for rotating turbulence produced by an oscillating grid
\cite{HBG82}, for decaying turbulence \cite{SDD08, MMRS11, D11},
  forced turbulence \cite{GCCM14}, and turbulent
  convection~\cite{KCG10}. The appearance of these large-scale
  vortices is associated to a noticeable two-dimensionalization of the
  flow in the plane perpendicular to the rotation axis. Rotating
  turbulent dynamics with Rossby number $O(1)$ is typical of many
  industrial and geophysical applications, but key fundamental
  questions are still open. These are mostly connected to the nature
  of the interaction between the two-dimensional vortical structures
  and the underlying fully three-dimensional anisotropic turbulent
  fluctuations, and to the way this impacts the Lagrangian dynamics of
  particles dispersed in the flow.  In this paper, we empirically
  assess the Eulerian and Lagrangian statistical properties of
  rotating flow by using high resolution direct numerical simulations
  at unprecedented resolution. We present the first simultaneous
  study of Lagrangian and Eulerian properties, seeding the strongly
  rotating flow with billions of small-particles with and without
  inertia.  In particular, we investigate statistical events much
  larger than the root mean squared fluctuations, measuring high order
  moments of velocity increments both along the rotation axis and in
  the perpendicular plane. To disentangle the statistical properties
  of the 2D structures from the underlying 3D turbulent background, we
  propose to decompose the velocity field on its instantaneous mean
  profile, obtained by averaging along the rotation axis, and on the
  fluctuations around it. We show that there exits a highly
  non-trivial entanglement among the vortical structures and the 3D
  background leading to a complex non-Gaussian distribution for both
  2D and 3D components. Similarly, we quantify the singular role
  played by vortical structures for the preferential concentration of
  inertial particles' trajectories.  We assess for the first time the
  properties of inertia in driving light and heavy particles advected
  by the rotating flow assessing the relative importance of the
  Centrifugal, Coriolis, added mass and Stokes forces and we show that
  rotation is extremely efficient in separating heavy from light
  particles, defeating the mixing properties of the underlying
  turbulent flow.

{\sc Eulerian fields}.  Rotation causes the generation of inertial
waves in the flow \cite{G68}. Waves, and the associated instabilities,
are of general interest given their fundamental character in
atmospheric and oceanographic applications. The interplay between
inertial waves and the two-dimensional three-components (2D3C)
turbulent structures that develop in rotating turbulent flows is the
subject of an active debate. Several
authors~\cite{CMG97,SW99,CRG04,BGSC06,S14,G03} have discussed the
possibility of describing the dynamics of rapidly rotating 3D flows
(limit of Rossby number much smaller than 1), in terms of wave
turbulence triggered by triadic resonant interactions (for reviews on
wave turbulence see, Refs. \cite{ZLF92,Na11,Ne11}). At the same time,
experimental \cite{BPSS03,YVS13} and numerical studies
\cite{CCEH05,SMRP12} indicate that 2D turbulence provides an effective
description of many aspects of rotating flows (for a recent review on
2D turbulence see~\cite{BE12}).\\ Theoretical
studies~\cite{CRG04,BGSC06}, addressing inertial wave turbulence
theory with a complete numerical solution in addition to the results
of quasi-normal closures, and numerical simulations\cite{SW99} have
shown that the nonlinear wave interactions tend to concentrate energy
in the wave-plane normal to the rotation axis, favoring the transfer
of energy from the 3D fast modes toward the 2D slow manifold (see also
\cite{CMG97} for a generalized quasi-normal approach, not restricted
to the asymptotic limit and with quantitative comparisons to direct
numerical simulation data). This has been proposed as a mechanism
that creates the columnar vortices~\cite{GM15}.  In particular,
triadic wave interactions are able to capture the main part of the so
called ``spectral buffer layer'', i.e., the spectral region close to
the 2D slow manifold \cite{BGSC06}.  On the other hand, the leading
resonant three-wave interactions cannot transfer energy directly to
the 2D modes \cite{SW99,CCEH05} and the wave approximation cannot be
uniform as a function of the wavenumber. In other words, wave
turbulence description ceases to be valid for very small wavenumbers
in the direction of the rotation axis $k_{||} = {\bm k}\cdot {\bm
  \Omega}/{\Omega} \simeq 0$ and for very large wavenumbers in the
perpendicular direction $k_{\perp} = {\bm k} - {\bm k}\cdot {\bm
  \Omega}/{\Omega}$ \cite{G03}, see also \cite{CRG04} for a discussion
about the decoupling of the 2D manifold. In such spectral regions the
coupling of modes by near-resonant and non-resonant triads has been
numerically investigated at moderate Rossby numbers by
\cite{CLM16}. Previously, the decoupling of the 2D slow mode was
questioned in~\cite{CRG04}, while a theoretical work \cite{G15} based
on stability analysis of the 2D flow has shown the existence of a
critical Rossby number below which 3D
rotating flow becomes exactly 2D in the long-time limit.\\ The
scenario is complicated by the fact that the predictions obtained from
the wave turbulence in infinite domains, with a continuous wavenumber
space, could differ from the observations of numerical simulations and
experiments, which necessarily deal with fluids confined in finite
volumes. Note, in particular, that the exact decoupling of the 2D slow
manifold from the inertial waves, due to resonant three-wave
interactions, is not proven in the continuous case (see
\cite{CRG04}). The discretization of the wavenumbers in finite volumes
causes a gap between the 2D manifold and the 3D modes that  could
favor the decoupling of the 2D dynamics (see Ref.  \cite{SW99} and
references therein).  The wave turbulence theory has been recently
applied to the case of an infinite fluid layer confined between two
solid boundaries~\cite{S14}.  In this case, the discretization of the
$k_{||}$ allows to address the dynamics of the 2D manifold and its
relationship with the wave-modes.  In particular, it has been shown
that the presence of a strong 2D mode might have a strong feedback on
the waves dynamics, as inertial waves can be scattered by the
vortices~\cite{S14}.  Along this line, recent experiments~\cite{YS14}
and numerical simulations~\cite{CCMDM14} has shown that a significant
fraction of the kinetic energy is concentrated in the inertial waves
whose period is shorter than the turnover time of the 2D structures,
while waves with longer period are scrambled by the turbulent
advection. Finally, recent numerical investigation of the rotating
Taylor-Green flow\cite{Al15} have shown that the limits of small
Rossby and large Reynolds numbers do not commute, and could lead to
different asymptotic regimes, displaying either the wave-turbulence or
the quasi-2D inverse cascade. As a result, the combined information
from theory, numerics and experiments is still far from being
sufficient to make a clear picture of the rotating turbulence. It is
safe to say that we do not control the physics of rotating turbulence
for realistic setup, in the presence of confinement, with external
forcing and at Rossby number $O(1)$, concerning both mean spectral
quantities and fluctuations on top of them.
\begin{figure}
\centering
\includegraphics[width=1.0\columnwidth]{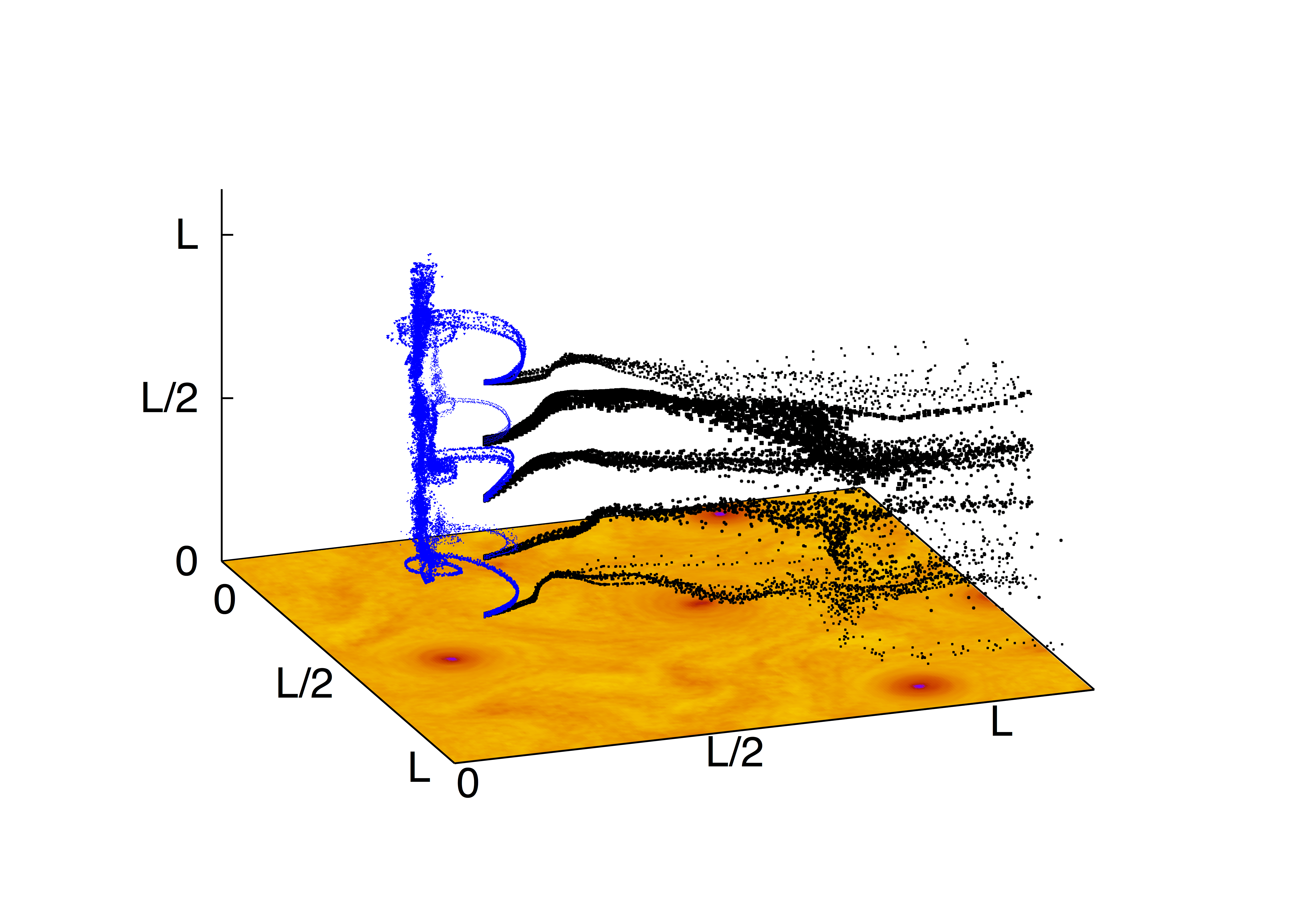}
\includegraphics[width=1.0\columnwidth]{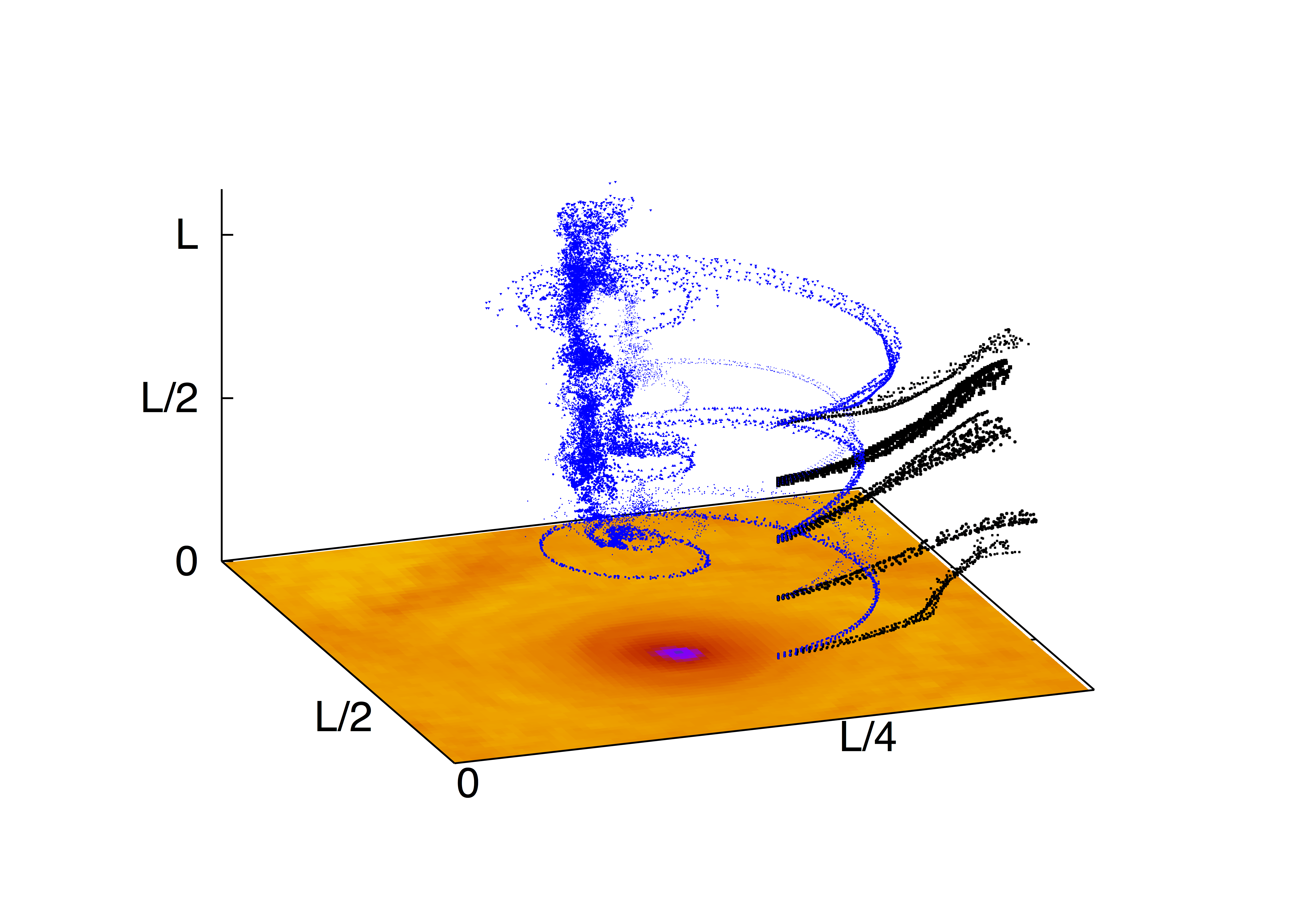}
\caption{3D rendering of the evolution of two different puffs of
  particles, one light (blue) and one heavy (black),
  released in a turbulent flow at Rossby number $Ro =0.25$. Particles
  are injected on the same rotation axis and with a velocity equal to
  that of the underlying fluid. The dispersion dynamics follows two
  different evolutions: light particles get trapped by the
  nearest columnar vortex and diffuse mainly vertically, while heavy
  particles tend to avoid the columnar structures and diffuse mainly
  horizontally. In the bottom plane, we show the intensity of the
  vertical vorticity averaged along the rotation axis. Bottom panel shows an enlargment of the top panel close to 
one intense vertical structure.}
\label{fig:2}
\end{figure}

\noindent 
{\sc Lagrangian particles}.  Lagrangian dynamics in rotating flows is
at the core of many different physical and engineering problems,
ranging from the dispersion and diffusion of pollutants, living
species or mixing of chemical reagents, to cite just a few
examples. However, the bulk of knowledge collected about the Eulerian
properties of the flow has no counterpart in the Lagrangian
framework. In the past decade a significant advance in the
understanding of the dynamics of inertial particles suspended in
turbulent flows has been achieved, notably for homogeneous and
isotropic flows \cite{TB2009}.  In the specific framework of particle
dynamics in rotating flows, very few results are available.  We
mention a theoretical prediction for the spatial distribution of
small, heavy particles in rotating turbulence ~\cite{EKR98}, a
prediction for the dispersion of fluid tracers in rotating
turbulence~\cite{CGNV04}, and the experimental study of the
tracer-like particles acceleration statistics ~\cite{DC11}.

In this paper, we present the first attempt to assess the importance
of Coriolis and centrifugal forces on the dynamical evolution and
spatial dispersion of light and heavy particles, within the
point-particles approximation. We show that the combined effect of
inertia plus rotation leads to a singular behavior for the particles'
statistics. In particular, the preferential sampling of high or low
vorticity regions is strongly enhanced and characterized by
anisotropic contributions on opposite directions: light particles tend
to diffuse mainly vertically (i.e. along the rotation axis) while
heavy particles are strongly confined in horizontal planes (see
Fig. \ref{fig:2}). As a result, the relative importance of Coriolis,
centrifugal or added-mass forces might vary by orders of magnitude
comparing light or heavy families.  We suggest that, at any rotation
rate of practical interest, both the 2D and the 3D turbulent
structures are coupled together and that any attempt to separate them
into a weak wave turbulence coupled with a quasi-2D slow dynamics in
the plane perpendicular to the rotation axis might fail to capture key
properties for both Eulerian and Lagrangian statistics.  This is an
important remark for the phenomenology of Eulerian and Lagrangian
rotating turbulence and to further improve its modelization.

This paper is organized as follows. In section (\ref{sec:dns}) we
discuss the numerical setup concerning both Eulerian and Lagrangian
properties. In section (\ref{sec:eulerian}) we discuss the Eulerian
statistical properties at changing both Rossby and Reynolds numbers,
while in section (\ref{sec:lagrangian}) we present the main results
concerning the dispersion of light or heavy particles. Our conclusions follow
in section (\ref{sec:conc}).

\section{NUMERICAL METHODS}
\label{sec:dns}

\subsection{Equation of motion for the Eulerian flow and for the Lagrangian trajectories}
The dynamics of an incompressible velocity field ${\bm u}$ in a
rotating reference frame with angular frequency ${\bm \Omega}$ is
given by the three dimensional Navier-Stokes equations (NSE):
\begin{equation}
\label{eq:navierstokes}
\frac {\partial \bm{u}}{\partial t} + \bm{u} \cdot \nabla \bm{u} +2{\bm \Omega} \times {\bm u}
= -\frac{\nabla p}{\rho_f} + \nu\Delta\bm{u} + \bm{f}. 
\end{equation} 
Here $\rho_f$ and $\nu$ are the density and the kinematic viscosity of
the fluid, respectively; $2{\bm \Omega} \times {\bm u}$ is the
Coriolis force, and $\bm{f} $ is an external force. For an
incompressible fluid, rotation breaks the statistical isotropy of the
flow, but not its homogeneity. Note that the centrifugal force ${\bm
  \Omega} \times {\bm \Omega} \times (\bm{r}-\bm{r}_0)$, which depends
on the distance from the position of the rotation axis, $\bm{r_0}$, is
absorbed in the pressure $p$, which is determined by the
incompressibility condition $\nabla \cdot \bm{u} = 0$.
\begin{table*}
\begin{ruledtabular}
\begin{tabular}{|cccccccccccccccc|}
$N$& $\Omega$ & $k_{\Omega}$ & $\nu$ & $\epsilon$ & $\epsilon_f$& $ u^2_0$ &  $\eta/dx$& $\tau_\eta/dt$ & $Re_\lambda$ & $Ro$ & $f_0$ & $\tau_f$ & $T_0$ & $\alpha$ 
\\  \hline
 1024 & 4  & 7 & $7 \times 10^{-4}$ &  1.2   &   $1.2$    & 1.05 & 0.67 & 120& 150 & 0.78  & 0.02 & 0.023 & 0.17 & 0.0\\
 1024 & 10 & 48 & $6 \times 10^{-4}$ &  0.46  &   $0.59$   & 1.6 &  0.76 & 294 & 580 & 0.24  & 0.02 & 0.023  & 0.25 & 0.1\\
 2048 & 4  & 7 & $2.8\times 10^{-4}$& 1.2    &   $1.2$    & 1.05 & 0.67  & 380 & 230 & 0.76  & 0.02 & 0.023 & 0.17 & 0.0\\
 2048 & 10 & 48 & $2.2 \times 10^{-4}$& 0.45 & $0.64$    & 1.7 & 0.72  & 550 & 1170 & 0.25  & 0.02 & 0.023 & 0.3 & 0.1 \\
 4096 & 10 & 49 & $1 \times 10^{-4}$ & 0.46  & $0.65$    & 1.7 & 0.78  & 1010 & 1600 & 0.25  & 0.02 & 0.023 & 0.3 & 0.1\\
 \end{tabular}
\end{ruledtabular}
\label{table:1}
\caption{Eulerian dynamics parameters. $N$: number of collocation
  points per spatial direction; $\Omega$: rotation rate; 
      $k_{\Omega}$: the Zeman wavenumber; $\nu$: kinematic
  viscosity; $\epsilon = \nu \int d^3x \sum_{ij} (\nabla_i u_j)^2$:
  viscous energy dissipation; $\epsilon_f = \int d^3x \sum_i f_i u_i
  $: energy injection; $u^2_0=1/3 \int d^3x \sum_i u_i^2$; $\eta =
  (\nu^3/\epsilon)^{1/4}$: Kolmogorov dissipative scale; $dx= L_0/N$:
  numerical grid spacing; $L_0= 2\pi$: box size; $\tau_\eta =
  (\nu/\epsilon)^{1/2}$: Kolmogorov dissipative time;
  $Re_{\lambda}=(u_0 \lambda)/\nu$: Reynolds number based on the
  Taylor micro-scale; $\lambda = (15 \nu u_0^2/\epsilon)^{1/2}$:
  Taylor micro-scale; $Ro = (\epsilon_f k_f^2)^{1/3}/\Omega$:
  Rossby number defined in terms of the energy injection properties,
  where $k_f=5$ is the wavenumber where the forcing is acting;
  $f_0$: intensity of the Ornstein-Uhlenbeck forcing; $\tau_f$:
  decorrelation time of the forcing; $T_0 =u_0/L_0$: Eulerian
  large-scale eddy turn over time; $\alpha$: coefficient of the
  damping term $\alpha \Delta^{-1}{\bm u}$. The typical total duration
  for a production run at resolution $N=2048$ is $T_{tot} = 20$.}
\label{tab:1}
\end{table*}
The regime of the flow is determined by the Reynolds number,
$Re_\lambda (see Table I)$, and by the Rossby number previously defined.  When
$Ro \gg 1$, the turbulent motions have time scales much shorter than
the rotation time-scale $\tau_{\Omega}$, and the flow is almost
unaffected by rotation. Rotation begins to affect the flow at $Ro
\sim O(1)$, when $\tau_{\Omega}$ is of the order of the
eddy-turnover-time at the forcing scale $1/(u_0k_f)$. A characteristic
scale of rotating turbulence is the Zeman wavenumber
\cite{HBG82,Z94,Delache} defined as the Fourier scale where the
inertial turnover time, $\tau_{nl}(k) = \varepsilon^{-1/3}k^{-2/3}$
becomes of the same order of $\tau_\Omega$, i.e., $k_\Omega\sim
(\Omega^3/\varepsilon)^{1/2}$, $\varepsilon$ being the energy transfer
rate. For $Ro \le 1$, the dynamics of the energy transfer will be
largely influenced by rotation. Importantly enough, as soon as the
Zeman wavenumber is larger than $k_f$, an inverse energy transfer
develops for $k \le k_{f}$, characterized by a strong accumulation of
the kinetic energy into $2D$ large-scale structures. As a result, for
$Ro \le 1$, the system develops a forward cascade of energy, partially
affected by the presence of rotation, and a simultaneous inverse
energy cascade leading to a strong anisotropy. The need to resolve
both interval of scales is the major bottleneck for direct numerical
simulations.\\ In the reference frame rotating with angular frequency
$\Omega$, the equations for the trajectory ${\bm r}_t$ and the
velocity ${\bm v}({\bf r}_t,t)$ of a small sphere of radius $R$ and
density $\rho_p$ suspended in the fluid field ${\bm u}$ can be
approximated as ~\cite{MR83}:
\begin{eqnarray}
\dot {\bm r}_t  &=& {\bm v}\,,
\label{eq:rotmaxeyriley1}
\\ 
\dot {\bm v} 
&=& 
\beta  D_t \bm{u}
-\frac{1}{\tau_p}(\bm{v}-\bm{u})
- 2 {\bm \Omega} \times ( \bm{v} -\beta \bm{u}) \nonumber\\
&&- (1-\beta) \left({\bm \Omega} \times ({\bm \Omega} \times ( \bm{r}_t -\bm{r}_0))\right)\,,
\label{eq:rotmaxeyriley2}
\end{eqnarray}
where $\bm{r}_0$ is the position of the rotation axis. Within the
point-particle model,the inertial dynamics is controlled by two
nondimensional parameters, the density ratio, $\beta = 3
\rho_f/(\rho_f+2\rho_p)$, and the Stokes number,
$St=\tau_p/\tau_\eta$, defined as the ratio between the particle
relaxation time, $\tau_p = R^2/3\beta\nu$, and the Kolmogorov time,
$\tau_\eta$. The first term on the rhs of
(\ref{eq:rotmaxeyriley2}) is the fluid acceleration and results from
an estimate of the added-mass and pressure gradients along the
trajectory of the tracers. The second term is the Stokes drag.  With
respect to the case of homogeneous and isotropic flows two new forces
appear in the rhs: the Coriolis and the centrifugal or centripetal, the
third and fourth terms respectively. To our knowledge, this is the
first attempt to assess the effects of these two forces on the
statistical and dynamical properties of inertial particles in
turbulence. At variance with the NSE for the flow, the centrifugal
force is present in the equation for the particle motion and it
explicitly breaks homogeneity because of its dependency on the
distance from the rotation axis. Its sign depends on the factor
$(\beta-1)$: for heavy particles ($0 \le \beta< 1$) the force is
centrifugal, while for light particles ($1 < \beta \le 3$) it is
centripetal.\\ In equation (\ref{eq:rotmaxeyriley2}), we have
neglect the Basset history and gravity forces, and the Faxen
corrections; moreover, we  approximate the material derivative
along the inertial particle trajectories in terms of the material
derivative along tracer paths. In the previous setup, tracer
trajectories are evolved according to the equation: $\dot {\bm r}_t =
{\bm u}({\bm r}_t,t)\,$.
\begin{table}
\begin{ruledtabular}
\begin{tabular}{|cccc|}
 Family & $\beta$& St   & type    \\ \hline
 T0 &  -       & - &  {\bf Tracer}       \\ \hline
 H1 & 0.4       &0.3 &  {\bf Heavy}      \\
 H2 & 0.4       &0.7 &         \\
 H3 & 0.8       &0.3 &         \\
 H4 & 0.8       &0.7 &         \\ \hline
 L5 & 1.2       &0.3 &  {\bf Light}       \\
 L6 & 1.2       &0.7 &         \\
 L7 & 1.6       &0.3 &       \\
 L8 & 1.6       &0.7 &       \\
 L9 & 1.6       &1  &          \\
 L10 & 1.6       &5 &         \\
 \end{tabular}
\end{ruledtabular}
\label{tab:2}
\caption{Lagrangian dynamics parameters. $\beta = 3
  \rho_f/(\rho_f+2\rho_p)$, ratio of the fluid and the particle
  densities; $St = \tau_p/\tau_\eta$: Stokes number. We evolve 10
  different families of inertial particles, plus a family of
  tracers. An ensemble of $N_a=5\times 10^5$ particles for each family
  is injected on 128 different rotation axis, located in different
  positions inside the simulation volume. Additionally, a set of
  $N_r=4 \times 10^6$ particles per family is uniformly injected in
  the flow, in order to optimize the statistical sampling of the whole
  simulation volume.}
\end{table}

\subsection{Direct Numerical Simulations setup}
We perform a set of state-of-the-art high-resolution direct
numerical simulations of the NSE in a periodic, cubic domain of size
$L=2 \pi$ with up to $N^3 = 4096^3$ collocation points. The rotation
axis is in the $x$-direction, i.e, ${\bm \Omega} = (\Omega,0,0)$. The
integration of Eqs. (\ref{eq:navierstokes}) has been performed by
means of a fully dealiased pseudo-spectral code, with the second-order
Adams-Bashforth scheme with viscous term exactly integrated. The
parameters of the Eulerian dynamics for the different runs are
reported in Table~\ref{tab:1}. The integration of
eqs. (\ref{eq:rotmaxeyriley1}) is performed by interpolating the
Eulerian velocity field and its derivatives with a $6-th$ order
B-spline algorithm on the particle position \cite{michel}. The parameters
of Lagrangian dynamics can be found in Table~\ref{tab:2}.
\begin{figure}
\centering
\includegraphics[angle=0,width=1.0\columnwidth]{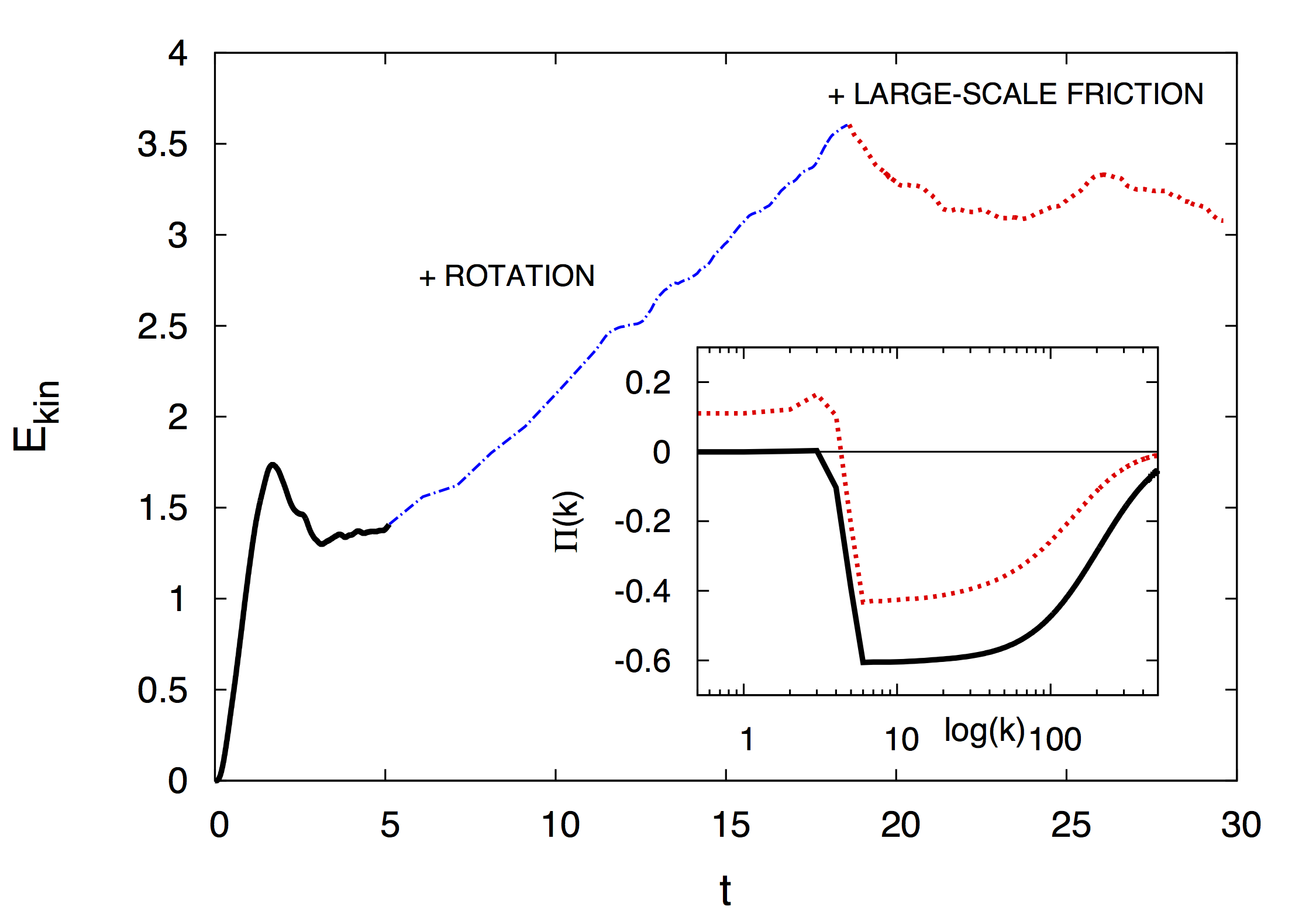}
\caption{(colors online) Kinetic energy evolution for a typical
   run with large rotation rate, in the presence of an
  inverse energy cascade. We show: the thermalization regime when
  rotation is not applied (black continuous line); the inverse cascade
  regime after rotation is switched on (blue dashed line); the
  stationary regime obtained by the application of a large scale
  friction (red dotted line). Inset: kinetic energy flux measured at
  the two stationary regimes: without rotation (black continuous
  line), and with rotation and large-scale friction (red dashed line),
  in the DNS with large rotation rate $\Omega=10$.}
\label{fig:energy}
\end{figure}
At high rotation, the presence of a simultaneous forward and inverse
cascade asks for an optimized setup, to minimize spurious finite-size
effects. The {\it critical} Rossby number where energy starts to flow
upscale is known or believed to depend on the way the system is forced
\cite{SMRP12,Al15,Da16} and on the aspect ratio of the volume where
the flow is confined \cite{DBLM14,SW99,S14}. In the presence of an inverse
flux, it is crucial to force the system at intermediate wavenumbers to
allow the large-scale flow to develop its own dynamics, without being
directly influenced by the forcing. Moreover, an energy sink mechanism
must be added to prevent the formation of a condensate at the lowest
Fourier mode that could spoil the statistics at all scales. \\ To
match the previous requirements, we adopt a stochastic isotropic
Gaussian force, ${\bm f}$, active on a narrow shell of wavenumbers at
$k_f \in [4:6]$. The  vector of each forcing Fourier mode is
obtained as ${\bf f}({\bf k},t) =f_0 (i\,{\bf k} \times {\bf X}
(t))$. The variables $X_i(t)$ are an independent, identically distributed
time-differentiable stochastic processes, solution of the following
Ornstein-Uhlenbeck 2-nd order process:
\begin{eqnarray}
&&  d X_i(t) = \\
  && -\left(\frac{1}{\tau_f} X_i(t) -\frac{1}{8 \tau_f^2}
\int_0^t X_i(t') dt'\right) dt + \sqrt{\frac{1}{4 \tau_f^3}} d W_i(t). \nonumber
\label{eq:fstochastic}
\end{eqnarray}
In the expression above, $\tau_f$ is the correlation time of the
process, and $W_i(t)$ is a Wiener process. It is important to stress
that the above time-correlated process ensures the continuity of the
Lagrangian acceleration of the tracers (see Ref. \cite{S91} for
details). At low Rossby number, to arrest the inverse cascade, we remove
energy at large scales with a linear friction term, $\alpha
\Delta^{-1}{\bm u}$ and acting on wavenumbers $|{\bm k}|\le 2$
only. This term is added to evolve the Lagrangian particles on a
stationary state without the need to overresolve the field in the
infrared regime.\\

To understand the basic phenomenology of a rotating turbulent flow, it
is useful to recall the different dynamical states that can be
observed in the case of strong rotation, i.e. low Rossby number. In
Figure~ (\ref{fig:energy}), we show the temporal evolution of the
total kinetic energy, $E_{kin} = \int d\bm{p}\, |{\bm u}_{{\bm
    p}}|^2$, starting from a fluid at rest and until a stationary
regime is achieved. In the early stage, the rotation rate $\Omega$ is
zero, and the flow develops a 3D direct cascade. Small-scale
thermalization is indicated by the overshoot of the kinetic
energy. This is the standard situation of stationary, non-rotating
turbulent flows where the energy input is balanced by viscous
dissipation. After this stage, we switch on the rotation and the
inverse energy cascade starts to develop if $\Omega$ is large enough:
this is indicated by the linear growth in time of the kinetic
energy. Later, we switch on the damping term at large scale.  Doing
that, we end up with a statistically stationary regime for a strongly
rotating turbulent flow. \\ In the inset of Fig. 3  we show
the presence of a simultaneous positive and negative spectral flux
when the rotation rate is large enough, indicating the existence of a
forward and inverse energy transfer at  scales smaller and larger of
the forcing scale, respectively.  We remark that the spectral flux is
here defined as the transfer of energy across a wavenumber $k$ by the
non-linear interactions $\bm{N}_{{\bf p}}$ of the Navier-Stokes
equations \cite{Po00}: $\Pi(k) = \int_{|{\bm p}|<k} d \bm{p} \,
\bm{u}^*_{{\bf p}}\cdot \bm{N}_{{\bf p}}$. The simultaneous presence
of direct and inverse cascades is shown by the two plateaus in the
spectral flux, in agreement with previous
findings~\cite{SCW96,EM98,SW99,CCEH05,BB07,MP09,DBLM14}.\\ Once the
turbulent flow is stationary, we seed it with Lagrangian particles of
different inertia, released with the same velocity of the underlying
fluid.
When $Ro$ is small, the flow is characterized by the presence of few
intense columnar cyclones, corotating with ${\bm \Omega}$. In the
plane perpendicular to the rotation, the associated two-dimensional
vortices are much slower than any other structure in the
flow. Moreover, as shown in Figure (\ref{fig:2}), these cyclones
strongly influence the distribution of the particles. Light particles
are trapped inside, while heavy particles are ejected, leading to an
extreme, singular preferential sampling of the underlying flow (see
sec. \ref{sec:lagrangian}). In all cases  investigate here, the flow
displays a few big cyclones well separated from each other. The
breaking of the cyclone-anticyclone symmetry is a well-known feature
of rotating turbulence \cite{Ba94,Na15,Al15,St67,Go01,Ge01}. It is
also the indication that the formation of the vortical columnar
structures cannot be entirely due to a 2D inverse cascade regime,
because in the plane perpendicular to the rotation axis the symmetry
is not broken. Nevertheless, it is suggestive to interpret the
presence of three long-living coherent columns in terms of the
dynamics of point vortices, since a system of three equal-sign
point-vortices is linearly stable in two dimensions
\cite{Aref2009}. We cannot exclude that the columnar vortices would
eventually merge into a single cyclone, after long enough
time. Considering that vortices with equal sign repel each other, and
that their merging would cause the generation of an intense shear
between them, it is arguable that the process of a collapse is
unlikely to occur.

\section{Eulerian Statistics}
\label{sec:eulerian}
\subsection{Fourier analysis}
\label{sec:fourier}
Rotation affects the spectral distribution of the kinetic energy on a
wide range of scales. For Fourier modes between the forcing and the
Zeman wavenumbers, $ k_f < k < k_\Omega$, a standard phenomenological
argument predicts for the energy spectrum:
\begin{equation}
E(k) = \int_{|{\bm p}| = k} d{\bm p} \langle |\bm{u}_{\bm p}|^2
\rangle \sim (\Omega \varepsilon)^{1/2} k^{-2}\,,
\label{eq:spectrum}
\end{equation}
which is obtained by estimating the typical transfer time in terms of
the non-linear time and  the rotation time, $\tau_{tr}(k) \propto
\tau_{nl}(k)^2/\tau_\Omega(k)$ \cite{Z95,MZ96,CHA2007}, see
Ref. \cite{BGSC06} for possible phenomenological extensions that also  take
into account anisotropic contributions.
\begin{figure}
\centering
\includegraphics[width=1.0\columnwidth]{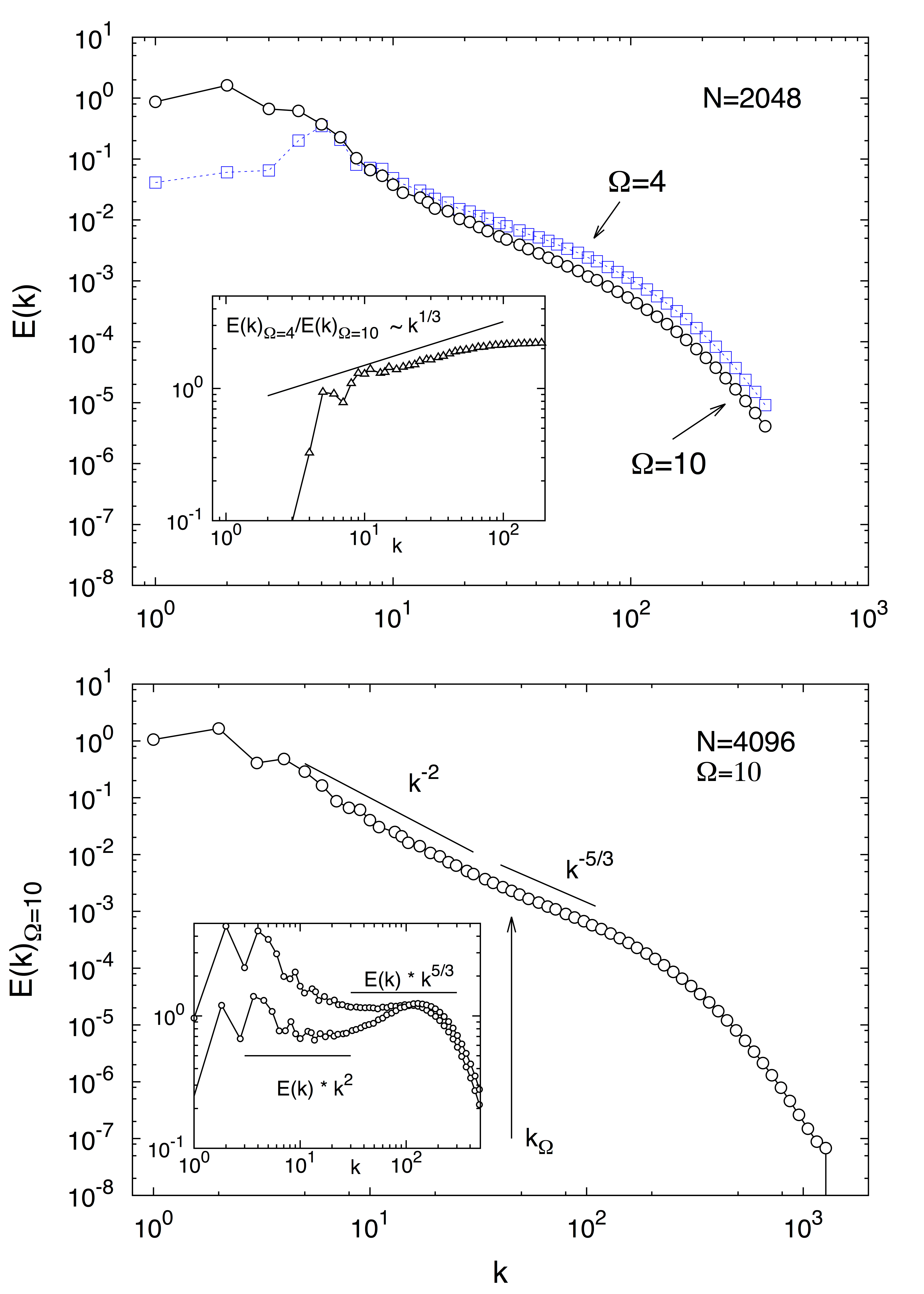}
\caption{Log-log plots of the energy spectrum. Top: Ppectra for
  the runs at $N=2048$. Data with $Ro=0.76$ and $\Omega=4$ (squares) and 
  data with $Ro=0.25$ and $\Omega=10$ (circles). Inset: Effect of
  rotation on the forward energy cascade is highlighted in terms of
  the ratio of the two spectra $E_{\Omega=4}(k)/E_{\Omega=10}(k) \sim
  k^{1/3}$. Bottom: Spectrum for $N=4096$, $Ro=0.25$ and
  $\Omega=10$; the expected scaling behaviors $\propto k^{-2}$ and
  $\propto k^{-5/3}$, above and below the Zeman wavenumber
  $k_{\Omega}$ respectively, are also plotted. Inset: Spectrum for
  $N=4096$ and $\Omega=10$ compensated with $k^{-2}$, and with
  $k^{-5/3}$. }
\label{fig:spectra}
\end{figure}
For small Rossby, $\Omega=10$, and at small wavenumbers $k < k_f$, we
observe the development of an inverse energy transfer: evidence is
given in the top panel of Fig.~\ref{fig:spectra}, where we compare
two spectra at low and high Rossby numbers, for the cases of
resolution $N^3=2048^3$. At larger wavenumbers, $k > k_f$, rotation
causes a steepening of the energy spectrum in good qualitative
agreement with the prediction Eq. (\ref{eq:spectrum}). Note that when
    rotation is strong, $\Omega=10$, the computed Zeman wavenumber is
    $k_{\Omega}\simeq 48$, indicating that presumably rotation has
    weaker effects at very large wavenumbers towards the dissipative
    range.\\ On the other hand, for large values of the Rossby
number, $\Omega=4$ and $k_{\Omega}\simeq 7$, the classical
Kolmogorov scaling $E(k) \sim \varepsilon^{2/3} k^{-5/3}$ associated
with the direct cascade is observed, and no backward energy transfer for
$ k < k_f$ develops.
\begin{figure*}
\includegraphics[width=2.\columnwidth]{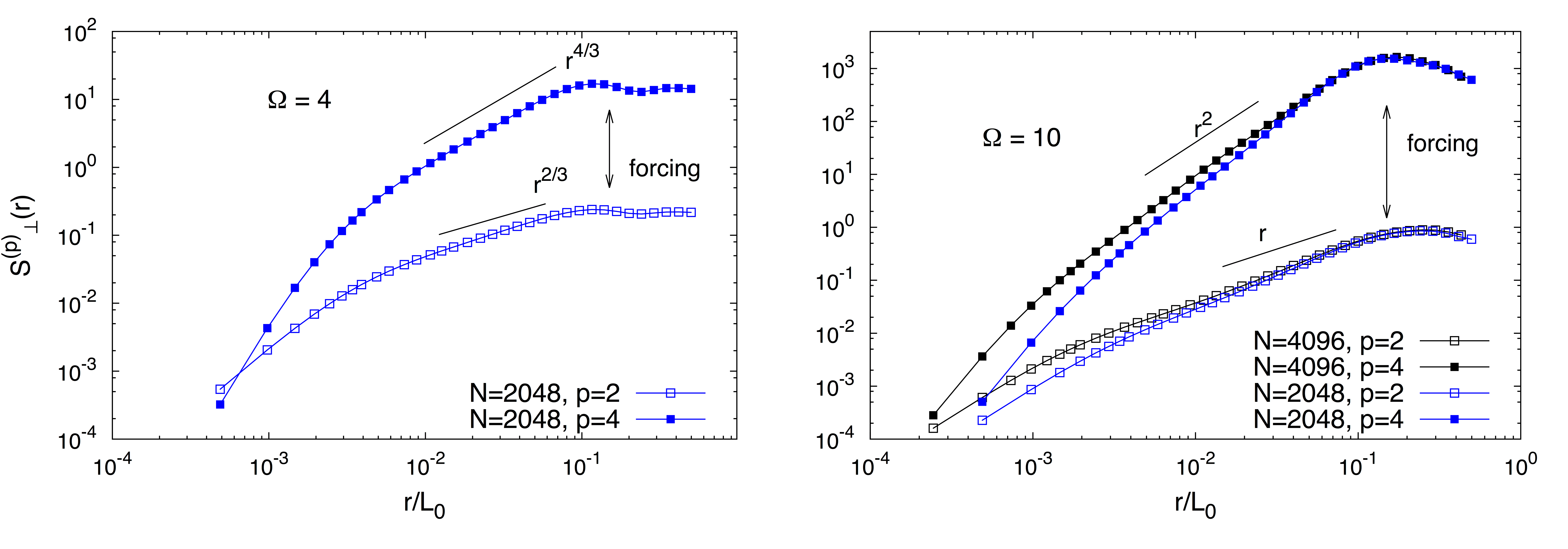}
\caption{Log-log plot of the 2nd, $S^{(2)}_{\perp}(r)$, and 4th order,
  $S^{(4)}_{\perp}(r)$, Eulerian transverse structure functions. Left:
  Case with $N=2048$ and $\Omega=4$. The dimensional scaling
  predictions $\propto r^{p/3}$ according to the K41 isotropic scaling
  are also plotted. Right: Case with $N=4096, 2048$ and $\Omega=10$.
  The dimensional scaling prediction $\propto r^{p/2}$ is also
  plotted.}
\label{fig:s2s4}
\end{figure*}
Remarkably enough, the change in the spectral exponent that takes
place by varying the Rossby number can be better identified by
plotting the ratio of two spectra, $E_{\Omega=4}(k)/E_{\Omega=10}(k)
\propto k^{-5/3}/k^{-2}$: when this is done, a clear $\simeq k^{1/3}$
behavior is observed (see the inset of Figure~\ref{fig:spectra}). In
the bottom panel of the  figure, we show the energy spectrum at
$N^3=4096^3$ resolution for the small Rossby number regime,
$\Omega=10$. The resolution is now sufficient to detect the transition
from the $k^{-2}$ to $k^{-5/3}$ scaling around the Zeman wavenumber,
as can be better appreciated in terms of the compensated plots in the
inset of Fig. 4. These results are in agreement with previous
numerical findings~\cite{MT07,Mi12}. \\ The
analysis in terms of the spectral properties cannnot be considered
conclusive. Spectra are not sensitive
to the Fourier phases, and, therefore, they are unable to distinguish if
the two-point spatial correlation is the result of a stochastic
turbulent background, or the result of coherent structures. Moreover, in
the presence of different physical scaling ranges (inverse cascade for
$k<k_f$, direct cascade plus rotation for $k_f <k < k_\Omega$ and
direct cascade with Kolmogorov phenomenology for $k>k_\Omega$), it is
impossible to detect power laws as a function of the wavenumber, even
at the highest resolution ever achieved as shown here. Finally, and
 more importantly, in order to assess the relative importance of
coherent and background fluctuations, it is mandatory to move to the
real space analysis, such as to have a direct way to assess
intermittency and deviations from Gaussian statistics {\it
  scale-by-scale} and for high-order velocity correlations.

\subsection{Real-space analysis}
A fundamental issue of rotating turbulence is to find suitable
observables that can disentangle the coupling between the 2D3C slow
modes and the 3D fast modes. A natural expectation is that the
strongest effects of the columnar vortices might manifest in the
statistics of the increments of the velocity components perpendicular
to ${\bm \Omega}$: $\delta u(r)_{\perp} = [({\bm u}({\bm x}+{\bm
    r})-{\bm u} ({\bm x})]\cdot {\hat {\bm t}}$, where the distance
${\bm r}$ is in the plane normal to ${\bm \Omega}$, and the versor
${\hat {\bm t}}$ is orthogonal to both ${\bm \Omega}$ and ${\bm
  r}$. Thus, we define the $p$-th order transverse structure function as
\begin{equation}
  \label{eq:transverse}
S^{(p)}_\perp(r) = \langle (\delta u(r)_{\perp})^p \rangle,
\end{equation}
where isotropy is assumed in the normal plane. In
Fig. (\ref{fig:s2s4}), we show the $2$-nd and the $4$-th order transverse structure function for
runs at different Reynolds and Rossby numbers. The scaling behaviors
indicate the existence of two different regimes in the inertial range
of scales, $ \eta < r < \ell_f$.
\begin{figure}
\centering 
\includegraphics[width=1.0\columnwidth]{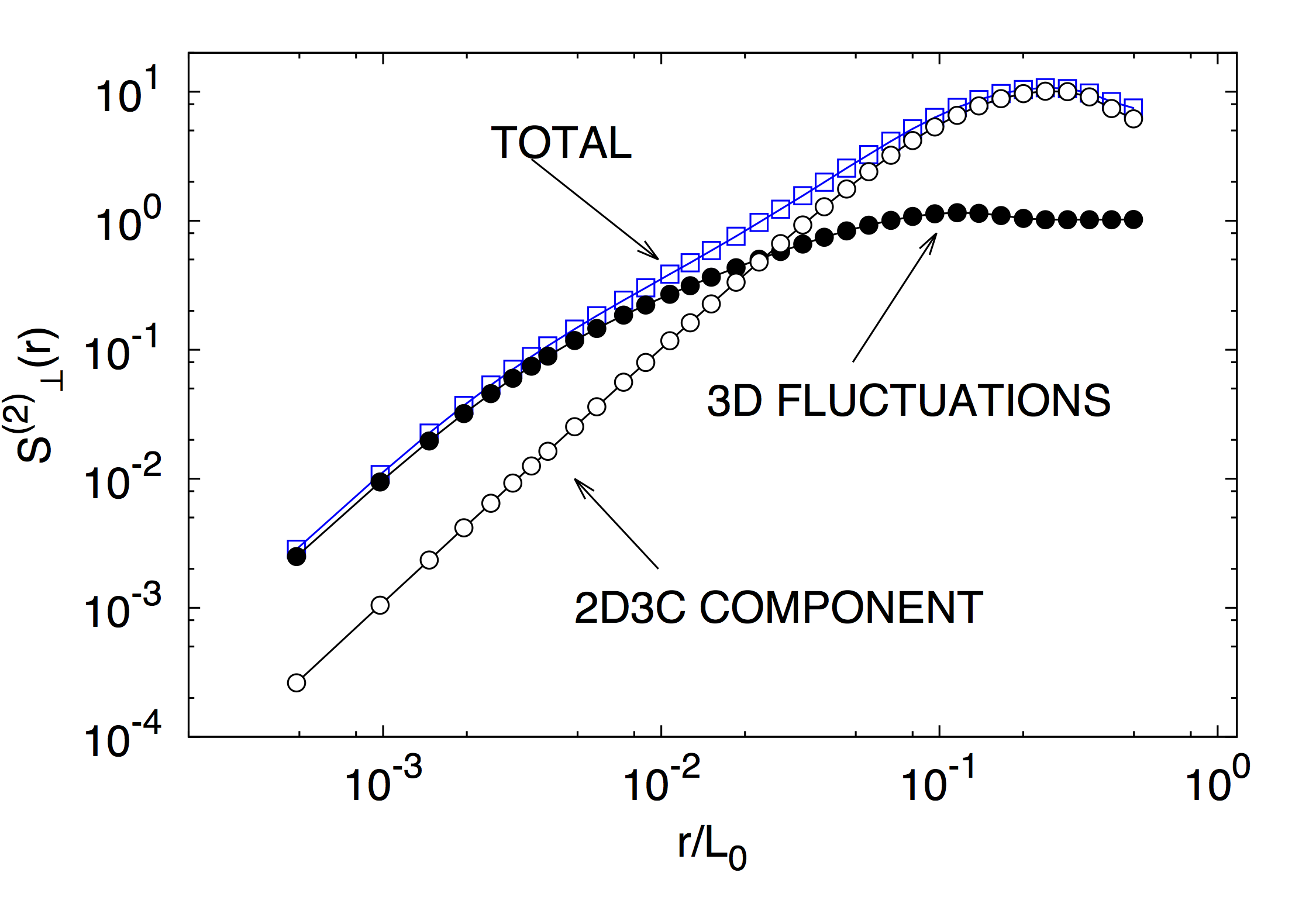}
\caption{log-log plot of the second order Eulerian transverse
  structure function in the plane perpendicular to the rotation axis,
  $S^{(2)}_{\perp}(r)$, for $N=2048$ and $\Omega=10$.  Whole field
  ${\bm u}(x,y,z)$ (squares); to the 2D3C ${\bm u}_{2D}(y,z)$
  component (empty circles), and to the fluctuating 3D component ${\bm
    u}'(x,y,z)$ (filled circles).}.
\label{fig:s2}
\end{figure}
In the right-hand panel, we show data at small $Ro$ ($\Omega=10$), and for
two different Reynolds numbers. A qualitative agreement with the
dimensional scaling $\propto r^{p/2}$ corresponding to the Zeman
phenomenology \cite{Po10,G03} is observed at large scale, while small
scales depend on the  Reynolds number and display a change in the local
slope by approaching the viscous scale at the highest resolution. \\
\begin{figure*}
\centering
\includegraphics[width=2.0\columnwidth]{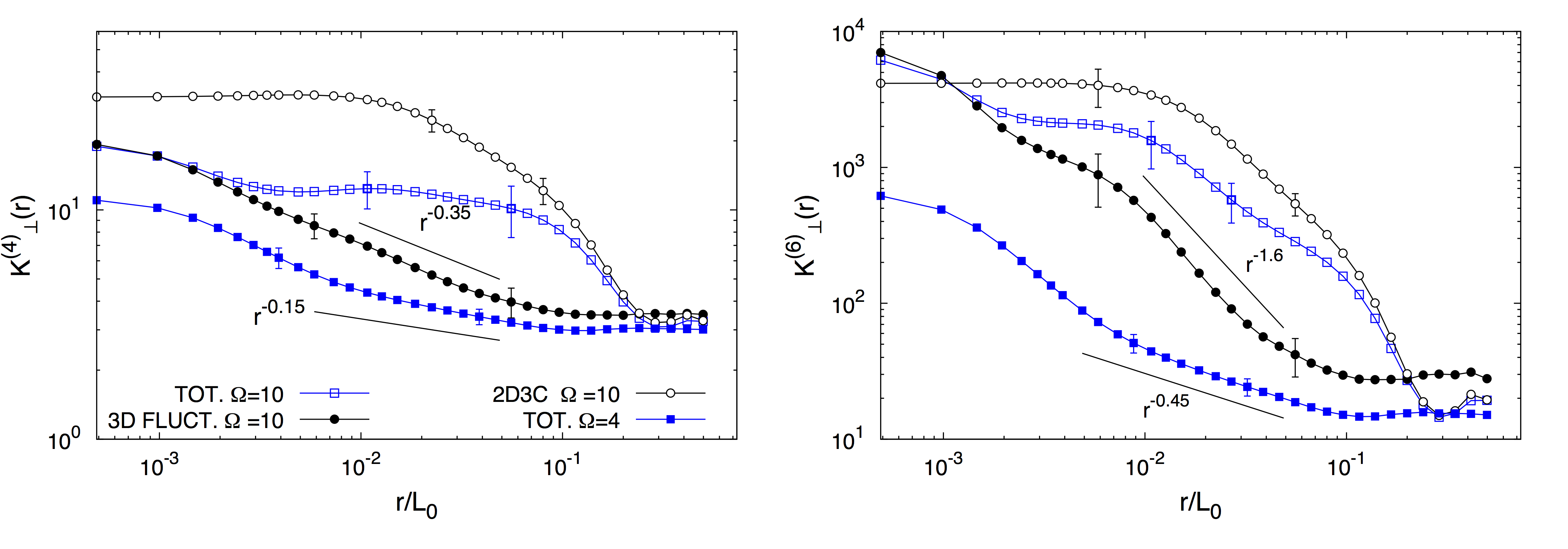}
\caption{Left: Log-log plot of the 4th-order flatness,
  $K^{(4)}_{\perp}(r)$, derived from the Eulerian structure functions
  transverse to the rotation axis, for data with $N=2048$. Data from
  DNS with large rotation rate, in the presence of an inverse cascade
  ($\Omega=10, Ro=0.25$): full field ${\bm u}$ (empty squares);
  2D3C ${\bm u}_{2D}$ component (empty circles); fluctuating 3D
  component ${\bm u}'$ (filled circles). Data from DNS with the
  direct cascade only, and no-columnar vortices ($\Omega=4, Ro=0.76$): full
  field ${\bm u}(x,y,z)$ (filled squares). We also superpose the power
  law prediction $\propto r^{-0.15}$ obtained from independent
  measurements of isotropic turbulence without rotation and the best
  fit for the power law measured with intense rotation in the present
  data $\propto r^{-0.35}$. Right: The same for the 6th-order
  flatness, $K^{(6)}_{\perp}(r)$ (same symbols). Error
  bars are estimated from different velocity snapshots and shown for a representative subset of points.}
\label{fig:flatness}
\end{figure*}
At high Rossby number ($\Omega=4$, left-hand panel of Fig. 5),
rotation effects are always subleading. Here, the statistics is in
good agreement with the Kolmogorov K41 prediction \cite{Po00}.  At $Ro
<<1$, the scaling laws are always spoiled by anisotropy and the only
systematic way to disentangle scaling properties would be to resort to
a decomposition in terms of eigenfunctions of the group of rotations
\cite{BP05,SO2SO3}. Moreover, the flow is naturally bimodal, with a
2D3C dynamics superposed and entangled with the 3D turbulent
fluctuations.\\ To better clarify the statistics {\it scale-by-scale},
we propose to decompose the velocity field into two components, one
given by the 2D3C slow modes and the other associated to the 3D fast
modes,
\begin{equation}
{\bm u}(x,y,z|t) = {\bm u}_{2D}(y,z|t) + {\bm u}'(x,y,z|t)\,.
\label{eq:decomposition}
\end{equation} 
Here we have define the two-dimensional field as the average of the
velocity field in the direction of ${\bm \Omega}$: $ {\bm
  u}_{2D}(y,z|t) = \int dx\, {\bm u}(x,y,z|t)$. In Fig.
(\ref{fig:s2}), we plot the second-order transverse structure
function, $S^{(2)}_\perp(r)$ measured for the undecomposed field, and
for the two fields obtained by the above decomposition. The figure
shows the existence of a scale, of the order of $l_{\Omega} = 2
\pi/k_{\Omega}$, where the statistics changes from being 2D3C to 3D
dominated. The background field follows quite closely the Kolmogorov
scaling $\propto r^{2/3}$ (not shown), while the 2D3C field has a
scaling much smoother than the Zeman estimate. This does not
necessarily contradict the results of section
(\ref{sec:fourier}). Rather, it clearly shows that within the Eulerian
statistics there are two different components that influence the
physics at different scales. It also suggests that any attempt to
fit or predict scaling laws without a separation of the different
contributions might lead to uncontrolled approximations.

In Fig.(\ref{fig:flatness}), we plot the 4th-order (left panel) and
6th-order (right-hand panel) flatness derived from the transverse structure
functions,
$$K^{(4)}_{\perp}(r)\equiv
\frac{S^{(4)}_{\perp}(r)}{(S^{(2)}_{\perp}(r))^2}; \qquad
K^{(6)}_{\perp}(r)\equiv
\frac{S^{(6)}_{\perp}(r)}{(S^{(2)}_{\perp}(r))^3}$$
for the undecomposed velocity field, the 2D projection, and the
fluctuating part. Except for very large spatial increments, the curves
are always far from the Gaussian limit. Consider the data for the
whole field at large rotation rate, i.e., $\Omega=10$ (empty squares
in both panels). The 4th-order flatness displays a weak dependence on
the analyzed scale in the inertial range, while the 6th-order does
change for scales smaller than the forcing range. How much are the
observed deviations from a Gaussian behavior  due to the presence
of the vortical columnar structures, and how much are they due to the 3D
turbulent fluctuations?\\ If we consider separately the statistics of
the 2D3C component (empty circles), ${\bm u}_{2D}$, and the statistics  of the 3D
fluctuations (filled circles), ${\bm u}'$, we find a surprising
result. The 4th-order flatness of the fast modes exhibits a strong
scale dependence. A scale-dependent flatness is the signature of
intermittency: here we observe it for both the 3D rapidly fluctuating
velocity field, and, to a smaller extent, the 2D3C slowly varying
component. The same trend is observed for the 6th-order flatness.
These results reveal that the reduction of intermittency previously
reported from data at smaller resolution and without a {\it
  scale-by-scale} analysis \cite{BPSS03,MT07,SMM08,MAP09} is merely
apparent and probably due to a nontrivial combination of effects
induced by the coherent structures and contributions from the
underlying 3D turbulent fluctuations. This is one of the main results
of this paper.\\ Finally, we note that naively one would expect
that the flatness of the fluctuating field for $\Omega=10$ should be
equal to the flatness of the total field for $\Omega=4$ (filled
squares). Our data show that this is not the case, meaning that
rotation not only leads to the formation of the 2D columnar
structures, but also modifies the 3D fluctuating turbulence, if the
Rossby number is small enough. We summarize in Table III the results for the
best fit to the flatness scaling exponents for the fluctuating
components at high rotation and for the total component at small
rotation rates.
\begin{table}
\begin{ruledtabular}
\begin{tabular}{|ccc|}
             & $\Omega = 4$  &  $\Omega=10$ \\ \hline
$\zeta(4)$   &  -0.15 (2)   & - 0.35(5)       \\ 
$\zeta(6)$   &  -0.45 (5)   & - 1.6(1)       \\ 
 \end{tabular}
\end{ruledtabular}
\label{tab:3}
\caption{Best fit to the scaling exponents of the p-th order Flatness,
  $K^{(p)}_{\perp}(r) \propto r^{\zeta(p)}$, with $p=4,6$.  For the
  high rotation $\Omega=10$ we fit the scaling for the fluctuating
  part only (filled circles in Fig. \ref{fig:flatness}).  For the case
  at low rotation rate $\Omega=4$ we fit the data for the whole
  undecomposed field because it coincides with the fluctuations (no
  vortical structures). Error refers to the uncertainty in the fit by
  changing the fitting scaling range.}
\end{table}

\begin{figure*}
\centering
\includegraphics[width=2.0\columnwidth]{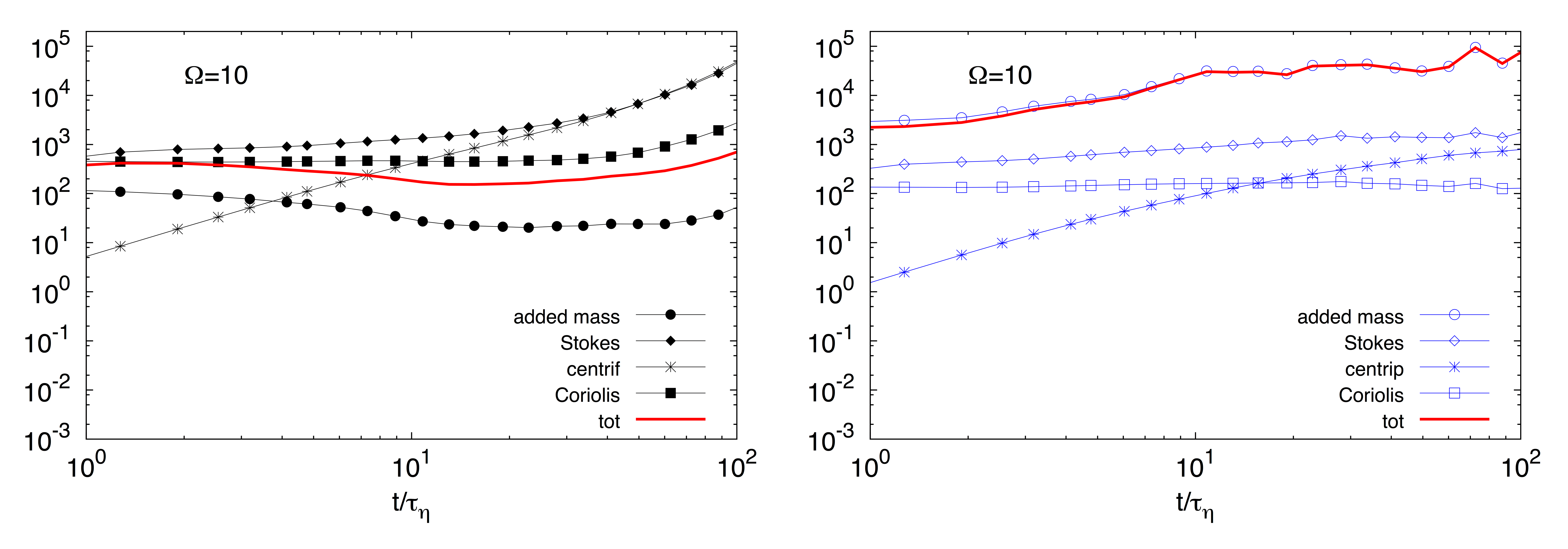}
\caption{Log-log plot of the time evolution of the contribution of the
  different forces to the rms particle acceleration, at resolution
  $N=2048$ and for high rotation (low Rossby number).  Left:  Inertial
  heavy particles, with $\beta=0.4$ and $St=0.7$ (family $H2$ in table
  II). Right: inertial light
  particles, with $\beta=1.6$ and $St=0.7$ (family $L8$ in table II).}
\label{fig:forces}
\end{figure*}

\section{Lagrangian Statistics}
\label{sec:lagrangian}
A novel way to investigate the statistics of the columnar vortices is
to exploit the peculiar features of the inertial particles in sampling
the flow. It is known that light particles are attracted inside the
vortices, while heavy particles are expelled out of them
\cite{Maxey1987,TB2009}. By studying the velocity statistics measured
along the trajectories of particles with different inertia, it is
possible to use their preferential concentration in specific flow
regions, to enhance or deplete the contribution of the slow vortical
modes with respect to the turbulent background.\\
\begin{figure}
\centering
\includegraphics[width=1.0\columnwidth]{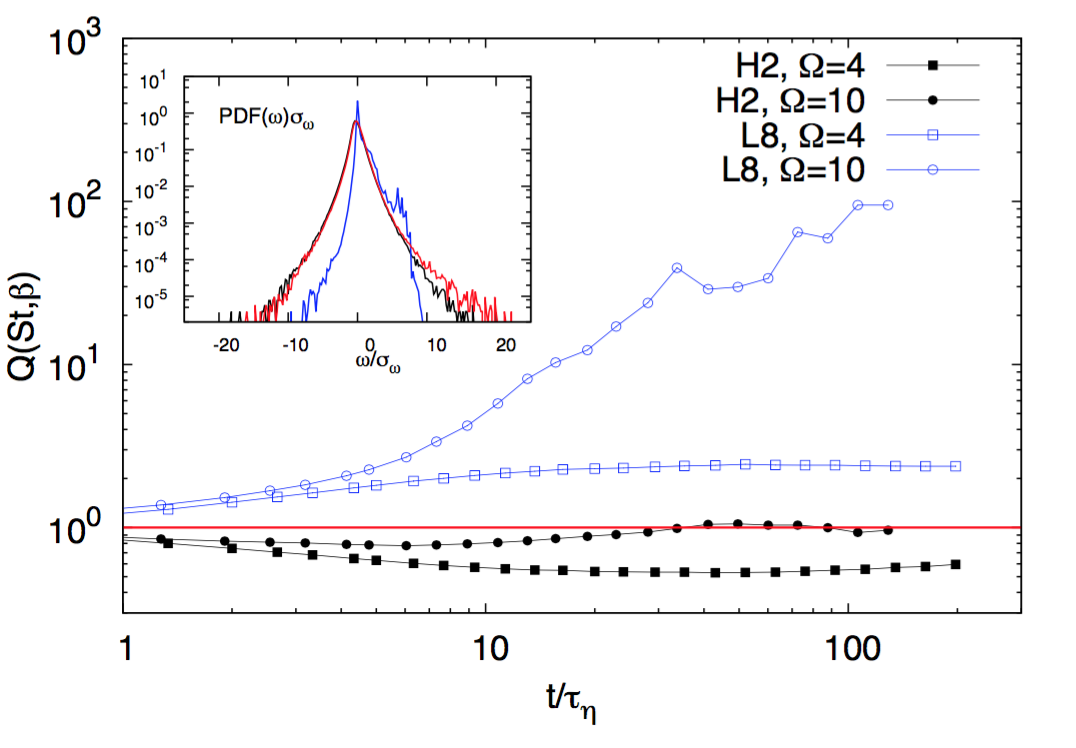}
\caption{A measure of the inertial particles preferential sampling of
  the vorticity regions at low, $\Omega=4$, and high, $\Omega=10$,
  rotation rates for two different families: a heavy one H$2$, and a
  light one L$8$. Inset: for the case with $\Omega=10$, the
  probability density function of the vertical vorticity, $\omega_x$,
  normalized to its standard deviation,as measured at the positions of
  light particles L$8$ (blue), heavy particle H$2$ (black) and tracers
  T$0$ (red). Data refer to DNS at resolution $N=2048$.}
\label{fig:lagr_vorticity}
\end{figure}
Here, we start by analyzing the different contributions of the forces
that influence inertial particles motion. In Fig.
(\ref{fig:forces}) we plot the time evolution of the
root-mean-squared  values of all accelerations:
\begin{equation}
\begin{cases}
a^{tot}_{rms}(t) = \langle \dot{\bm v}^2 \rangle; \qquad \text{total}\\
a^{am}_{rms}(t) = \beta^2 \langle (D_t {\bf   u})^2\rangle; \qquad \text{added mass}\\
a^{St}_{rms}(t) = 1/\tau_p^2\langle({\bf v} - {\bf u})^2\rangle; \qquad \text{Stokes drag} \\
a^{Co}_{rms}(t) = 4 \langle [{\bf \Omega} \times ({\bf v} - \beta
  {\bf u})]^2\rangle; \qquad \text{Coriolis}\\
a^{Cp}_{rms}(t) = (1-\beta)^2 \langle [{\bf \Omega} \times ( {\bf
    \Omega} \times ({\bf r}_t -{\bf r}_0)]^2\rangle; \text{centripetal}.\\
\end{cases}
\end{equation}
When 
the Rossby number is small, i.e. for $\Omega=10$,
the inertial particle dynamics does not always attain a statistically
steady state. The relative importance of the forces is 
affected for  two different reasons. The first one is purely kinematic,
since both Coriolis and centripetal forces are proportional to the
rotation rate. The second one is dynamical: the organization of the
flow, with the formation of strong columnar vortices, competes with
the kinematic effects. \\
\begin{figure*}
\centering 
\includegraphics[width=2.0\columnwidth]{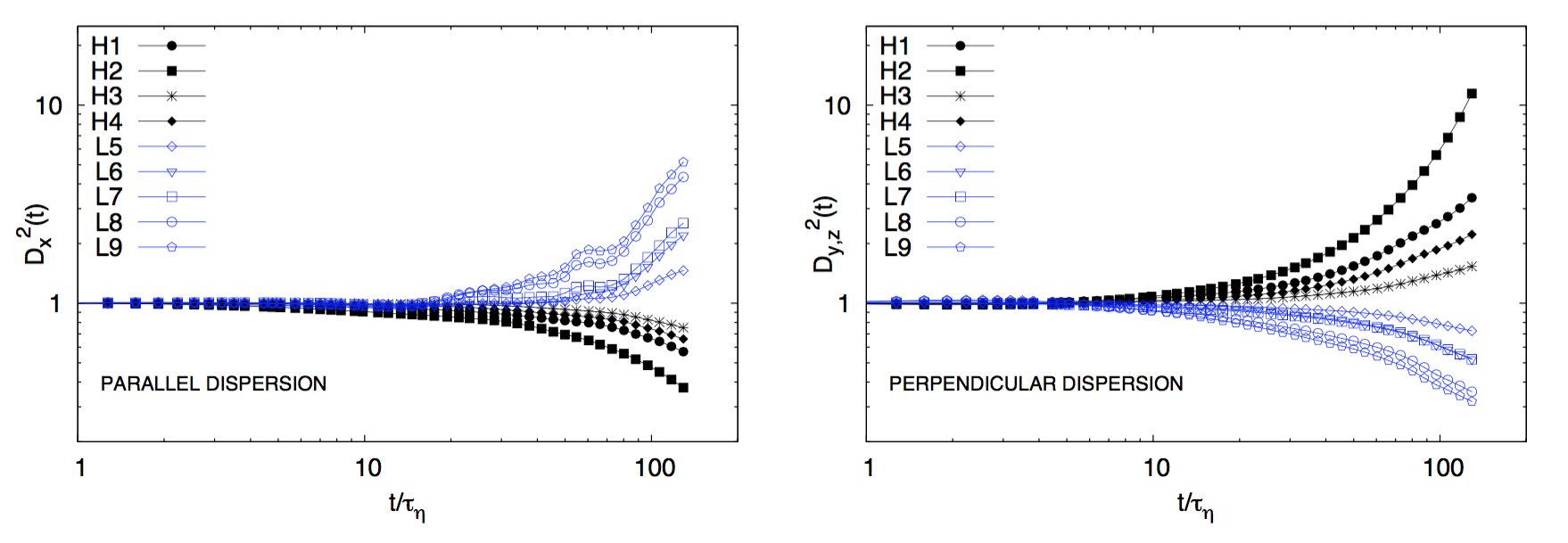}
\caption{Absolute dispersion of inertial particles in the direction
  parallel (left) and perpendicular (right) to the rotation axis. The
  mean square displacement of inertial particles is normalized with
  that of tracers. Labels refer to the four families of heavy
  particles H$1-4$ and to the 5 families of light particles
  L$5-9$. See table II for details. Data refer to the DNS with
  $N=2048$ and $\Omega=10$. }
\label{fig:6}
\end{figure*}
If particles are heavier than the fluid, the centrifugal force soon
becomes dominant: particles not only tend to avoid coherent vortical
structures, but also tend to spiral away from their rotation axis very
efficiently (see also Fig. \ref{fig:2}). This enhanced centrifugal action is balanced by the Stokes
drag only. Comparing, $a_{rms}^{St}$ and $a_{rms}^{Cp}$, it is
interesting to note that this balance is very efficient, leading to a
total acceleration, $a_{rms}^{tot}$, much smaller than the single
contributions, and to an almost stationary statistics in the 
long-time limit. In this regime, the dynamics of the heavy
particles is uncorrelated with respect of the underlying fluid. 
Particles move away from their rotation axis with a
spiral motion, whose radius grows exponentially in time, $r(t) \sim
\exp(\Omega^2 \tau_p t)$. Since their velocity also increases
exponentially over time, the particle Reynolds number might eventually
become too large for the validity of the model equations
[Eqs. (\ref{eq:rotmaxeyriley1}) and (\ref{eq:rotmaxeyriley2})]
\cite{MR83}. Hence, it would be crucial to perform a systematic
comparison with experimental data, in order to understand the
limitation of the pointlike approach in the limit of very heavy
particles.\\ For small Rossby number, we observe an opposite behavior
in the case of light particles. The centripetal force attracts the
light particles toward their original rotation axis, but its intensity
vanishes as $r_t \to r_0$. The overall effect is, therefore, to
spatially confine the trajectories of light particles, depleting
turbulent diffusion and preventing them from exploring regions far
away from the rotation axis. Additionally, one needs to consider the
dynamical attraction inside the coherent vortical structures. As
visually shown in Fig. \ref{fig:2}, preferential centripetal concentration is
the leading effect and almost all light particles are trapped inside
vortical structures. The leading term in light particles acceleration
is the added-mass, which is not balanced by other forces. We also
notice that at long times the temporal behavior of the added mass term
becomes noisy, in spite of the large number of particles used in computing the 
average. This is because eventually all light particles collapse into
the cores of a few columnar vortices, thus reducing the effective statistics.
\\ The singular role played by
the presence of the coherent structures for Lagrangian statistics is
better quantified in Fig. \ref{fig:lagr_vorticity}, where we plot
the preferential sampling of specific flow regions, by measuring the
average vertical vorticity at the particle positions normalized with
the averaged vertical vorticity in the volume,
\begin{equation}
Q_{St,\beta}(|t) = \frac{\langle [w_x({\bf
      r}_t,t)]^2\rangle_{\beta,St}}{\langle [w_x({\bf
      r}_t,t)]^2\rangle_{tracer}}\,.
\end{equation} 
At low rotation, $\Omega=4$, the preferential sampling by heavy or
light particles is similar to that observed in homogeneous and
isotropic turbulence, and quantitatively it is an $O(1)$ effect with
respect to the mean-fluid vorticity. At large rotation rate, the
situation is different: heavy particles, because of the sweeping due
to the centrifugal forces, do not show preferential concentration,
while light particles oversample the intense vorticity regions with
an effect that is a factor $O(100)$ larger. In
Fig.~\ref{fig:lagr_vorticity}, we also show the probability
distribution function (PDF) of the vertical vorticity $w_x$ along
particle trajectories for tracers and for one light and one heavy
family.  Notice the bimodal PDF for the light particles induced by the
trapping in the vortex cores; for the heavy particles, the PDF is
symmetric, because of the homogeneous sampling of the flow regions
outside strong vortical structures. \\ Concerning absolute dispersion,
the influence of the strong vortical structures will induce a
systematic anisotropic effect for tracers \cite{CGNV04}. On the
other hand, since the vortical structures are fatal traps for the
light particles and strong repellers for heavy particles, we expect to
measure strong deviations in the single-particle dispersion too. In
Fig. \ref{fig:6}, we show the mean-square absolute dispersion of
the particles from their initial position as a function of time
\begin{equation}
D^i_{St,\beta}(t) = \frac{\langle (r^i_t- r^i_0)^2 \rangle_{St,\beta}}{\langle (r^i_t-
  r^i_0)^2 \rangle_{tracer}}\,,
\label{eq:absolutedisp}
\end{equation}
  along different directions $i=(x,y,z)$, here normalized  with the ones
  measured for tracers. For the heavy particles, we find that the
  diffusion in the plane normal to the rotation axis ${\bm \Omega}$ is
  enhanced, due to the centrifugal effect, while parallel diffusion is
  reduced. Moreover, at a fixed value of the density mismatch $\beta$,
  the effect is stronger for higher Stokes number. \\ The diffusion
  behaviors are inverted for light particles. The trapping in the
  vortices strongly suppresses the transverse diffusing, but enhances
  the one parallel to the rotation axis (see also Fig. \ref{fig:2}).
  Because of the two dimensionalization induced by rotation, all the
  components of the fluid velocity are weakly dependent on the
  coordinate along the rotation axis. This occurs also for the
  component of the velocity parallel to ${\bm \Omega}$.  As a result,
  the columnar vortices can have a uniform coherent velocity in the
  direction of ${\bm \Omega}$.  Light particles, once trapped in the
  columnar structures, are transported almost ballistically along
  their axis, as in an elevator.

\section{CONCLUSIONS}
\label{sec:conc}
Rotating turbulence is key for many industrial and geophysical
applications. In many empirical setups it is also key to control the
dispersion and advection of particles. Very little is known concerning
the combined Eulerian-Lagrangian properties, and a long-lasting debate
exists concerning the effects of confinement and forcing, and whether or not  they have
a singular footprint on the statistics.  We present the results
of a state-of-the-art direct numerical simulation study of Eulerian
and Lagrangian rotating turbulence at high and low Rossby numbers. To
our knowledge, this is the first attempt to study the evolution of
particles in rotating turbulence and in the presence of both direct and
inverse energy cascade. At high rotation rates, we show that the
Eulerian ensemble strongly deviates from a self-similar
normal-distributed statistics at changing the analyzed scale, with a
key influence of the coherent vortical columnar structures. By
removing from the velocity field the 2D3C component, obtained by
averaging over the vertical direction, we are able to assess
quantitatively the degree of intermittency present in the remaining 3D
fluctuations: in particular, we show that there exists a
nontrivial non-Gaussian contribution also in the background
fluctuations.Whether this result is specific to an intermediate range
of Rossby numbers and Eulerian intermittency might or not decrease in
the limit of very small Rossby number is a question that needs further
investigation. We simultaneously measure the Lagrangian
statistics, following millions of light or heavy and tracer particles
injected in different rotation axes inside the rotating volume. We
 show, for the first time, that an extreme preferential sampling
develops as soon as there exists coherent structures in the flow and
that this has a singular effect for the fate of heavy or light
particles. In particular, heavy particles diffuse more efficiently in
the plane perpendicular to the rotation axis, while light particles
tend to diffuse only vertically. The discovery of this {\it elevator}
effect might have important implications for industrial applications
and for the population dynamics of passive or active microswimmers in
the oceans. Tracking light particles is also key to highlighting the
breaking of cyclonic-anticyclonic symmetry, a property of any 3D
rotating fluid at Rossby numbers $~O(1)$. Many issues remain open. It
would be extremely interesting to understand the degree of
universality of the 2D3C statistics and of the remaining 3D
fluctuations at changing the forcing mechanisms, the large-scale
friction (and the confinement aspect ratio). It is also expected, but
not measured yet, that the vortical structures will strongly influence
the two-particles Richardson dispersion in rotating flow. Similarly,
it is not known how much Lagrangian velocity increments along
particles trajectories are eventually affected by rotation, a key
point to build up stochastic models for particles dispersion in
atmospheric and marine environments. A detailed study of
single-particle and two-particles (relative dispersion) diffusion
statistics is a  next step in  Lagrangian dynamics exploration. \\ It
is still an open question to understand how to match the results from
finite volume experiments and direct numerical simulations with the
predictions of unforced wave turbulence in infinite domains (see also
\cite{B08} for a discussion of discreteness and resolution
effects). Finite volume effects can be estimated by comparing the
typical distance traveled by the waves during the duration of the
simulation, estimated in terms of their group velocity. In our case,
for the highest rotation case this distance is pretty small, of the
order of $10\%$ of the total volume. Using state-of-the-art highly
resolved DNS, as done here, is crucial in order to reduce the spectral
gap with the horizontal plane also. Here, for the highest resolved
case, we have an excellent resolution of the buffer layer near $k_{||}
=0$, including wavenumbers with angle close to 0.04 deg with the
horizontal plane and thus reducing finite volume effects. However,
improving angular resolution to better capture the spectral buffer
layer also at small wavenumbers close the two-dimensional manifold is
an issue \cite{CRG04}: further numerical investigations e.g., in slab
geometries, permitting to obtain values of $k_{||} =0$ small enough
are desirable to shed further light on the problem of 2D-3D modes
coupling.\\ Other numerical approaches meant to understand the
importance of different triadic interactions in Fourier space, and to
further clarify the nature of the inverse cascade in purely rotating
turbulence, are possible. One important example is given by Ref. 
\cite{SL05} where reduced Navier-Stokes equations including only
near-resonant, nonresonant and near-two-dimensional triad
interactions are considered. These numerical approaches are restricted
to work on spectral space, with severe limitation in the number of
modes that can be considered.  \\
\begin{acknowledgments}
Simulation has been performed at CINECA, within the PRACE Grant No
Pra092256.  We acknowledge the European COST Action MP1305 ``Flowing
Matter'' and funding from the European Research Council under the
European Union's Seventh Framework Programme, AdG ERC Grant Agreement
No 339032. ASL acknowledges support from MIUR, within projects PESCA
SSD and RITMARE. This work is part of the research program of the
Foundation for Fundamental Research on Matter (FOM), which is part of
the Netherlands Organisation for Scientific Research (NWO).
\end{acknowledgments}

\end{document}